\documentstyle[12pt,twoside,fleqn,esppre,epsfig]{article}

\newcommand{\be}{\begin{eqnarray}}
\newcommand{\ba}{\begin{array}}
\newcommand{\ea}{\end{array}}
\newcommand{\ee}{\end{eqnarray}}

\newcommand{\dslash}{\partial \hskip -0.5em /}

\newcommand{\vslash}{v \hskip -0.5em /}

\newcommand{\bD}{{\bf D}}
\newcommand{\bDp}{{\bf D}^{(\pi)}}

\newcommand{\AmS}{{\protect\the\textfont2
  A\kern-.1667em\lower.5ex\hbox{M}\kern-.125emS}}

\title{Hadron Structure Functions within a Chiral Quark
       Model\thanks{Talk prepared for the QNP 2000 conference in 
       Adelaide Feb. 2000 and to appear in the proceedings. 
       Presentation prevented by United Airlines. 
       This work is supported in parts by funds
       provided by the U.S. Department of Energy (D.O.E.)
       under cooperative research agreements
       \#DF--FC02--94ER40818 and \#DE-FG0398ER41066 and by the
       Deutsche Forschungsgemeinschaft (DFG) under 
       contract We 1254/3-1.}}%

\author{Herbert Weigel\address{Center for Theoretical Physics\\
       Laboratory of Nuclear Science and Department of Physics\\
       Massachusetts Institute of Technology,
       Cambridge, Ma 02139, USA}%
        \thanks{Heisenberg Fellow}
       and 
       Leonard Gamberg\address{Department of Physics and Astronomy\\
       University of Oklahoma, 440 West Brooks, Norman, Ok 73019, USA}
}

\begin{document}
\maketitle

\begin{abstract}
We outline a consistent regularization procedure to
compute hadron structure functions within bosonized
chiral quark models. We impose the Pauli--Villars scheme, which 
reproduces the chiral anomaly, to regularize the 
bosonized action. We derive the Compton amplitude from this action 
and utilize the Bjorken limit to extract structure functions 
that are consistent with the scaling laws and sum rules 
of deep inelastic scattering.
\end{abstract}

\section{THE CHIRAL MODEL}
\vskip-0.2cm
The bosonized action of chiral quark models can be cast in the form
\be
{\cal A} [S,P]=-iN_C{\rm Tr}_{\textstyle\Lambda}{\rm log}\, 
\left[i\dslash-\left(S+i\gamma_5P\right)\right]
-\frac{1}{4G}\int d^4x\, {\rm tr}\, {\cal V}(S,P)\, .
\label{bosact}
\ee
Here ${\cal V}$ is a local potential respectively for scalar and 
pseudoscalar fields $S$ and $P$ which are matrices in flavor space.
For example, in the Nambu--Jona--Lasinio (NJL) model~\cite{Na61} one
has ${\cal V}=S^2+P^2+2{\hat m}_0(S+iP)$. From the gap--equation 
we obtain the VEV, $\langle S\rangle=m$ which parameterizes the 
dynamical chiral symmetry breaking. The regularization of the
quadratically divergent quark loop is indicated by the cut--off $\Lambda$. 
We adjust its value as well as the coupling constant $G$ and the 
current quark mass ${\hat m}_0$ to fit the phenomenological meson parameters 
$m_\pi$ and $f_\pi$, leaving only a single free parameter, the 
constituent quark mass, $m$. An essential feature of these models is that 
the derivative term in (\ref{bosact}) is formally identical to that of 
a non--interacting quark model. Hence the current operator is given as 
$J^\mu={\bar q}{\cal Q}\gamma^\mu q$, with ${\cal Q}$ being a flavor
matrix. We compute expectation values of currents by
introducing pertinent sources in the bosonized action and taking 
appropriate derivatives.

The major concern in regularizing the functional (\ref{bosact}) is
to maintain the chiral anomaly. We achieve this goal by splitting this 
functional into $\gamma_5$--even and odd pieces and only regularize the 
former. The $\gamma_5$--odd part turns out to be conditionally finite. 

Details of this presentation are published in \cite{We99}. 
For related work see refs \cite{We96,Di96,Wa98}.

\section{REGULARIZATION OF THE COMPTON TENSOR}
\vskip-0.2cm
DIS off hadrons is parameterized by 
the hadronic tensor $W^{\mu\nu}(q)$ with $q$ being the momentum 
transmitted to the hadron by the photon. $W^{\mu\nu}(q)$ is obtained from the 
hadron matrix element of the commutator 
$[J^\mu(\xi), J^\nu(0)]$. In bosonized quark models we find it 
convenient to start from the absorptive part of the forward virtual 
Compton amplitude 
\be
T^{\mu\nu}(q)=\int d^4\xi\, {\rm e}^{iq\cdot\xi}\,
\langle p,s|T\left(J^\mu(\xi) J^\nu(0)\right)|p,s\rangle
\quad {\rm and }\quad
W^{\mu\nu}(q)=\frac{1}{2\pi} \Im\, (T^{\mu\nu}(q)\, .
\label{comp1}
\ee
(We denote the momentum of the hadron by $p$ and eventually  
its spin by $s$.) The advantage is that the time--ordered product is 
unambiguously obtained from the regularized action 
\be
T\left(J^\mu(\xi) J^\nu(0)\right)=
\frac{\delta^2}{\delta v_\mu(\xi)\, \delta v_\nu(0)}\,
{\rm Tr}_\Lambda {\rm log}\,
\left[i\dslash-\left(S+i\gamma_5P\right)+{\cal Q}\,
\vslash\right]\Big|_{v_\mu=0}\, \, .
\label{tprod}
\ee
In order to extract the leading twist pieces of the structure 
functions, we study $W^{\mu\nu}(q)$ in the Bjorken limit: 
$q^2\to-\infty$ with $x=-q^2/p\cdot q$ fixed.

We now have to specify the regularization of the functional
trace in (\ref{tprod}). We define
\be
i \bD = i\dslash - \left(S+i\gamma_5P\right)
+\vslash{\cal Q} \quad {\rm and}\quad
i \bD_5 = - i\dslash - \left(S-i\gamma_5P\right)
-\vslash{\cal Q}
\label{defd}
\ee
and separate the functional trace
into (un--)regularized $\gamma_5$--even (odd) pieces,
\be
{\rm Tr}_\Lambda {\rm log}\,
\left[i\dslash-\left(S+i\gamma_5P\right)+{\cal Q}\,
\vslash\right] &=&-i\frac{N_C}{2}
\sum_{i=0}^2 c_i {\rm Tr}\, {\rm log}
\left[- \bD \bD_5 +\Lambda_i^2-i\epsilon\right]
\nonumber \\ &&\hspace{1cm}
-i\frac{N_C}{2}
{\rm Tr}\, {\rm log}
\left[-\bD \left(\bD_5\right)^{-1}-i\epsilon\right]\, .
\label{PVreg}
\ee
With the conditions
$ c_0=1\, ,\,\, \Lambda_0=0 ,\, \sum_{i=0}^2c_i=0\,
{\rm and}\, \sum_{i=0}^2c_i\Lambda_i^2=0$ the double 
Pauli--Villars regularization renders the functional 
in (\ref{tprod}) finite. 

\section{PION STRUCTURE FUNCTION}
\vskip-0.2cm
DIS off pions is characterized by a single structure function,
$F(x)$. For its computation we have to specify the pion matrix 
element in the Compton amplitude (\ref{comp1}). Whence we
introduce the pion field ${\vec\pi}$ via\footnote{The 
coupling $g$ and the constituent quark mass $m$ are related by
the pion decay constant. In the chiral limit the relation is linear
$m=g f_\pi$.}
\be
S+iP\gamma_5=m\, \left(U\right)^{\gamma_5} = m\, {\rm exp}
\left(i \frac{g}{m}\gamma_5\,{\vec\pi} \cdot {\vec\tau} \right)\, .
\label{chifield}
\ee
Expanding (\ref{PVreg},\ref{chifield}) to linear and quadratic order  
in $\vec{\pi}$ and $v_\mu$, respectively yields the proper result for
the anomalous decay $\pi^0\to\gamma\gamma$. In turn we obtain the 
Compton amplitude for virtual pion--photon scattering
by expanding (\ref{PVreg},\ref{chifield}) to second order in both, 
${\vec\pi}$ and $v_\mu$.
Due to the separation into $\bD$ and $\bD_5$ this calculation differs 
from the evaluation of the `handbag' diagram because isospin violating 
dimension--five operators emerge. Fortunately all isospin 
violating pieces cancel yielding 
\be
F(x)=\frac{5}{9} (4N_C g^2)
\frac{d}{dm_\pi^2}\left\{m_\pi^2
\sum_{i=0}^2 c_i\, \frac{d^4k}{(2\pi)^4i}\,
\left[-k^2-x(1-x)m_\pi^2+m^2+\Lambda_i^2-i\epsilon\right]^{-2}\right\}\, .
\label{pionf}
\ee
The cancellation of the isospin violating pieces is a feature
of the Bjorken limit: insertions of the pion field on the propagator
carrying the infinitely large photon momentum can be safely 
ignored. Furthermore this propagator can be taken to be the one 
for non--interacting massless fermions. This implies 
that the Pauli--Villars cut--offs can be omitted for this propagator
leading to the desired scaling behavior of the structure function.

\section{NUCLEON STRUCTURE FUNCTIONS}
\vskip-0.2cm
In the bosonized chiral quark model baryons emerge as solitons of 
the meson fields~\cite{Al96}. We parameterize the soliton by
\be
U(\vec{x},t)=A(t)
{\rm exp}\left(i\vec{\tau}\cdot\hat{r}\Theta(r)\right)A^\dagger(t)\, ,
\label{soliton}
\ee
with the chiral angle $\Theta(r)$ being determined from the 
stationary condition for constant~$A$. Subsequently we quantize 
the collective coordinates~$A$ to generate nucleon states.

As argued above we take the quark propagator with the infinite photon
momentum to be free and massless. Thus, it is sufficient to differentiate
\be
&&
\frac{N_C}{4i}\sum_{i=0}^2c_i
{\rm Tr}\,\left\{\left(-\bDp\bDp_5+\Lambda_i^2\right)^{-1}
\left[{\cal Q}^2\vslash\left(\dslash\right)^{-1}\vslash\bDp_5
-\bDp(\vslash\left(\dslash\right)^{-1}\vslash)_5
{\cal Q}^2\right]\right\}
\nonumber \\ &&\hspace{0.5cm}
+\frac{N_C}{4i}
{\rm Tr}\,\left\{\left(-\bDp\bDp_5\right)^{-1}
\left[{\cal Q}^2\vslash\left(\dslash\right)^{-1}\vslash\bDp_5
+\bDp(\vslash\left(\dslash\right)^{-1}\vslash)_5
{\cal Q}^2\right]\right\}\, ,
\label{simple}
\ee
with respect to the photon field $v_\mu$ as in eq (\ref{tprod}). 
We have introduced the $(\ldots)_5$ description
\be
\gamma_\mu\gamma_\rho\gamma_\nu
=S_{\mu\rho\nu\sigma}\gamma^\sigma
-i\epsilon_{\mu\rho\nu\sigma}\gamma^\sigma\gamma^5
\quad {\rm and} \quad
(\gamma_\mu\gamma_\rho\gamma_\nu)_5
=S_{\mu\rho\nu\sigma}\gamma^\sigma+
i\epsilon_{\mu\rho\nu\sigma}\gamma^\sigma\gamma^5
\label{defsign}
\ee
to account for the unconventional appearance of 
axial sources in $\bD_5$~\cite{We99}. Upon substituting (\ref{soliton})
into (\ref{simple}) and computing the functional trace, using a basis
of quark states obtained from the Dirac Hamiltonian in the
background of $U(\vec{x},t)$, we find analytical results for the structure 
functions. By construction their regularization is consistent with the 
chiral anomaly. We refer to \cite{We99} for detailed formulae and the 
explicit verification of sum rules. Here we simply report the important 
result that the structure function entering the Gottfried sum rule 
is related to the $\gamma_5$--odd piece of the action and hence does not 
undergo regularization. This is surprising because in the parton model 
this structure function differs from the one associated with the Adler 
sum rule only by the sign of the anti--quark distribution. The latter 
structure function, however, gets regularized, in agreement with the
quantization rules for the collecive coordinates. As we have
consistently implemented the regularization at the level of the bosonized 
action this demonstrates that in effective models these structure functions 
are quite different from constituent quark distributions and in their 
description one has to go beyond identifying degrees of freedom.

\section{NUMERICAL RESULTS FOR THE POLARIZED NUCLEON STRUCTURE FUNCTIONS}
\vskip-0.2cm
Unfortunately numerical results for the full structure functions, 
{\it i.e.} including the properly regularized vacuum piece are not yet 
available. However, we have verified that in the Pauli--Villars 
regularization the axial charges are saturated to 95\% or more by 
their valence quark contributions once the self--consistent soliton 
is substituted. This provides sufficient justification to adopt
the valence quark contribution to the polarized structure functions
as a reliable approximation \cite{We96}. In Fig.~1 we compare 
the model predictions for the linearly independent polarized structure 
functions to experimental data~\cite{Abe98}.
\begin{center}
\parbox[t]{15.3cm}{\sf Fig. 1: Model predictions for the 
polarized proton structure functions $xg_1$ (left panel)
and $xg_2$ (right panel). The curves labeled `RF' denote the
results as obtained from the valence quark contribution to
(\ref{simple}). These undergo a projection to the infinite
momentum frame `IMF' \cite{Ga98} and a leading order `LO' 
DGLAP evolution \cite{DGLAP}. Data are from SLAC--E143 \cite{Abe98}. }
\bigskip

\epsfig{figure=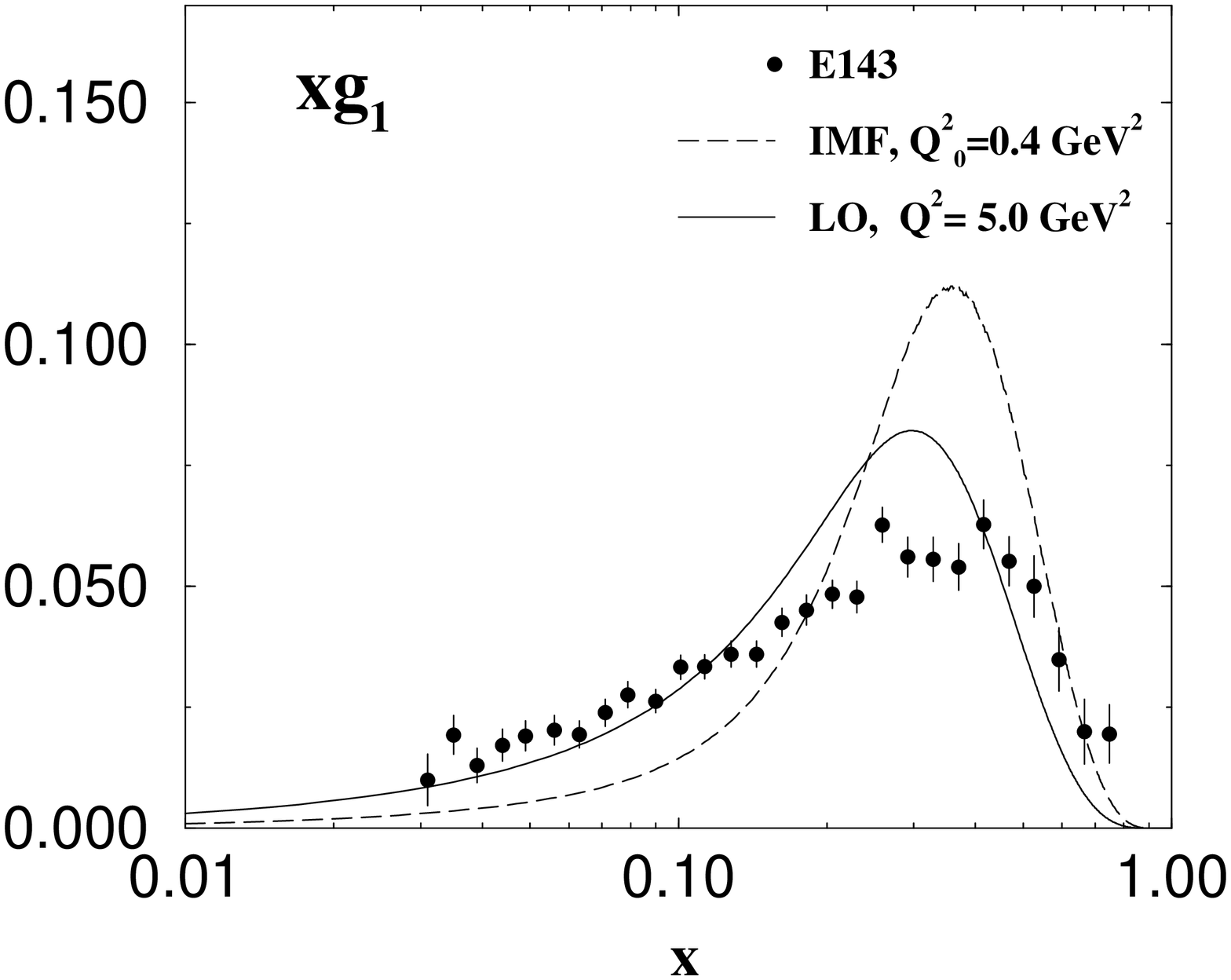,height=7.0cm,width=4.3cm,angle=270}
\hspace{1cm}
\epsfig{figure=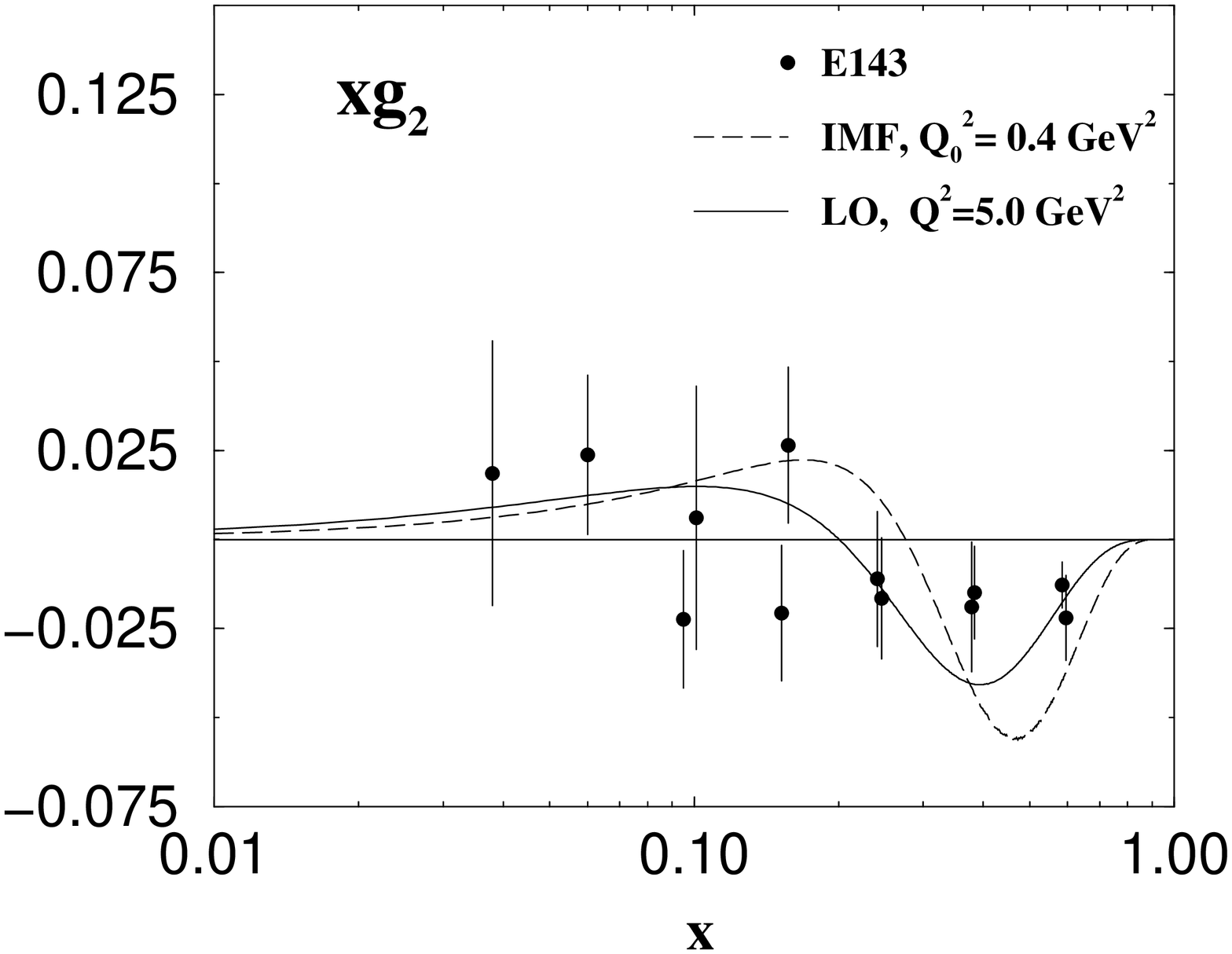,height=7.0cm,width=4.3cm,angle=270}
\end{center}
\smallskip
The evolution of the structure function $g_2$ to the momentum scale of 
the experiments requires the separation into twist--2 and twist--3 
components \cite{DGLAP}. We observe that the model results for the 
polarized structure functions, which we argued to have reliably approximated, 
agree reasonably well with the experimental data. This encourages future
work in this direction.

\section*{ACKNOWLEDGMENTS}
\vskip-0.2cm
We thank our collaborators E. Ruiz Arriola, O. Schr\"oder and
H. Reinhardt for valuable contributions. HW is particularly grateful to 
the organizers of the conference for including this contribution in 
the proceedings despite its presentation being prevented by United Airlines.


\begin{thebibliography}{9}
\small
\vskip-0.2cm
\bibitem{Na61}
Y. Nambu and G. Jona--Lasinio,
\newblock Phys. Rev. {\bf 122} (1961) 345; {\bf 124} (1961) 246.
\bibitem{We99}
H. Weigel, E. Ruiz Arriola and L. Gamberg,
Nucl. Phys. {\bf B560} (1999) 383.
\bibitem{We96}
H. Weigel, L. Gamberg and H. Reinhardt,
Mod. Phys. Lett. {\bf A11} (1996) 3021,\
Phys. Lett. {\bf B399} (1997) 287,\
Phys. Rev. {\bf D55} (1997) 6910;\
L. Gamberg, H. Reinhardt and H. Weigel,
Phys. Rev. {\bf D58} (1998) 054014;\
O. Schr\"oder, H. Reinhardt and H. Weigel,
Phys. Lett. {\bf B439} (1998) 398; \
H. Weigel, 
hep--ph/9902390.
\bibitem{Di96}
D. I. Diakonov {\it et al.},
Nucl. Phys. {\bf B480} (1996) 341,\
Phys. Rev. {\bf D56} (1997) 4069.
\bibitem{Wa98}
M. Wakamatsu and T. Kubota,
Phys. Rev. {\bf D57} (1998) 5755,\
Phys. Rev. {\bf D60} (1999) 034020.
\bibitem{Al96}
R. Alkofer, H. Reinhardt and H. Weigel,
\newblock Phys. Rep. {\bf 265} (1996) 139;\
C. V. Christov {\it et al.}, Prog. Part. Nucl. Phys. {\bf 37} (1996) 91.
\bibitem{Ga98}
L. Gamberg, H. Reinhardt and H. Weigel,
Int. J. Mod. Phys. {\bf A13} (1998) 5519.
\bibitem{DGLAP}
G. Altarelli, P. Nason and G. Parisi, Phys. Lett. {\bf B320} (1994) 152,
{\bf B325} (1994) 538 (E);\,
A. Ali, V.M. Braun and G. Hiller, Phys. Lett. {\bf B226} (1991) 117.
\bibitem{Abe98}
K.\ Abe {\it et al.},
Phys. Rev. {\bf D58} (1998) 112003.
\end{thebibliography}
\end{document}